# Direct Measurement of the Timescale of Spin Reorientation in $TbMn_6Sn_6$

Sinéad A. Ryan[1], Anya Grafov[1], Na Li[1*], Hans T. Nembach[2,3], Justin M. Shaw[4], Hari Bhandari[5], Tika R Kafle[1,6], Richa Sapkota[1], Henry C. Kapteyn[1,7], Nirmal J. Ghimire[5], Margaret M. Murnane[1]

[1]JILA, University of Colorado Boulder, 440 UCB, Boulder, CO 80309, USA

[2]Associate Physical Measurement Laboratory, National Institute of Standards and Technology, Boulder, CO 80305, USA

[3]Department of Physics, University of Colorado Boulder, Boulder, CO 80309, USA

[4]Physical Measurement Laboratory, National Institute of Standards and Technology, Boulder, CO 80305, USA

[5]Department of Physics and Astronomy and Stavropolous Center for Complex Quantum Matter, 438 Nieuwland Science Hall, University of Notre Dame, Notre Dame, IN 46556, USA

[6]Center for Integrated Nanotechnologies, Los Alamos National Laboratory, Los Alamos, NM 87545, USA

[7]KMLabs Inc., Boulder, CO 80301, USA

*Corresponding author: Na.li-1@colorado.edu

## ABSTRACT

Kagome ferrimagnets are of fundamental interest because of their tunable magnetic, electronic and topological properties, with potential applications for quantum-enabled technologies and spintronics. Here we present the first direct measurement of the intrinsic timescale of the spin reorientation transition in the Kagome ferrimagnet $TbMn_6Sn_6$. This material exhibits a magnetic phase transition near room temperature, where spins remain collinear while the total magnetic moment rotates from out-of-plane to in-plane. This reorientation has been attributed to the competing anisotropies of Tb and Mn, whose magnetic moments have very different temperature dependencies. By probing at the Mn *M*-edge using high harmonic extreme ultraviolet pulses, we measure reorientation timescales between 6 and 18 ps, depending on the laser excitation fluence. Our model of the magnetization dynamics shows that this timescale is consistent with a spin reorientation driven by very large anisotropy energies, on meV energy scales. Our model also predicts a possibility of a 180° reorientation of the out-of-plane moment, that could facilitate optically controlled magnetization switching between very stable ground states, for applications in spintronics or data storage.

## I. INTRODUCTION

The ability to control spin orientation and transport in magnetic materials is key to applications in information storage and spintronic technologies [1–4]. In particular, spin reorientation transitions, where the easy axis of magnetization switches between distinct crystallographic directions, provide a natural setting to study how competing anisotropies control collective spin behavior [5–7]. Such transitions not only reveal the fundamental physics of anisotropic competition, but also offer a route for switching between distinct magnetic states – a key requirement for functional spintronic devices [8]. As a result, there has been increasing interest in the field of ferrimagnetic spintronics [1], as well as ferrimagnetic spin reorientations, due to their potential application in stress-mediated magnetoelectric memory cells (MELRAMs) [2,3].

The Kagome ferrimagnet ($TbMn_6Sn_6$, Tb166) is a prototypical material system that exhibits a reorientation temperature $T_{\mathrm{SR}} \approx 309$ K and a ferrimagnetic Néel temperature of $T_{\mathrm{FN}} \approx 423$ K, above which the ferrimagnetic order vanishes (some literature equivalently refers to this as the Curie temperature). In this material, the transition metal (TM) Mn atoms form a Kagome net, and are ferrimagnetically coupled with rare-earth (RE) Tb atoms. This leads to two competing temperature-dependent anisotropy fields: Tb favors out-of-plane alignment, whereas Mn tends to align in-plane. As the temperature approaches $T_{\mathrm{SR}}$, the Tb moment and its associated uniaxial anisotropy weaken more rapidly than those of Mn, shifting the balance and driving a collective 90° rotation of all spins from out-of-plane to in-plane, while maintaining ferrimagnetic order. Figure 1(a) depicts the temperature-dependent spin-alignment in Tb166 in three specific temperature regimes ($T < T_{\mathrm{SR}}$, $T_{\mathrm{SR}} < T < T_{\mathrm{FN}}$, and $T > T_{\mathrm{FN}}$). Below $T_{\mathrm{SR}}$, the Mn and Tb moments are ferrimagnetically aligned along the out-of-plane axis. As the temperature increases above $T_{\mathrm{SR}}$, the moments rotate into the basal plane, where the in-plane Mn moment reaches its maximum value. Upon further heating toward the ferrimagnetic Néel temperature $T_{\mathrm{FN}}$, thermal fluctuations increase spin disorder, eventually leading to a fully randomized state above $T_{\mathrm{FN}}$. Static magnetization, transport, and neutron scattering studies have established this equilibrium picture [9–15], making Tb166 an archetype for studying spin reorientation in rare-earth transition-metal ferrimagnets.

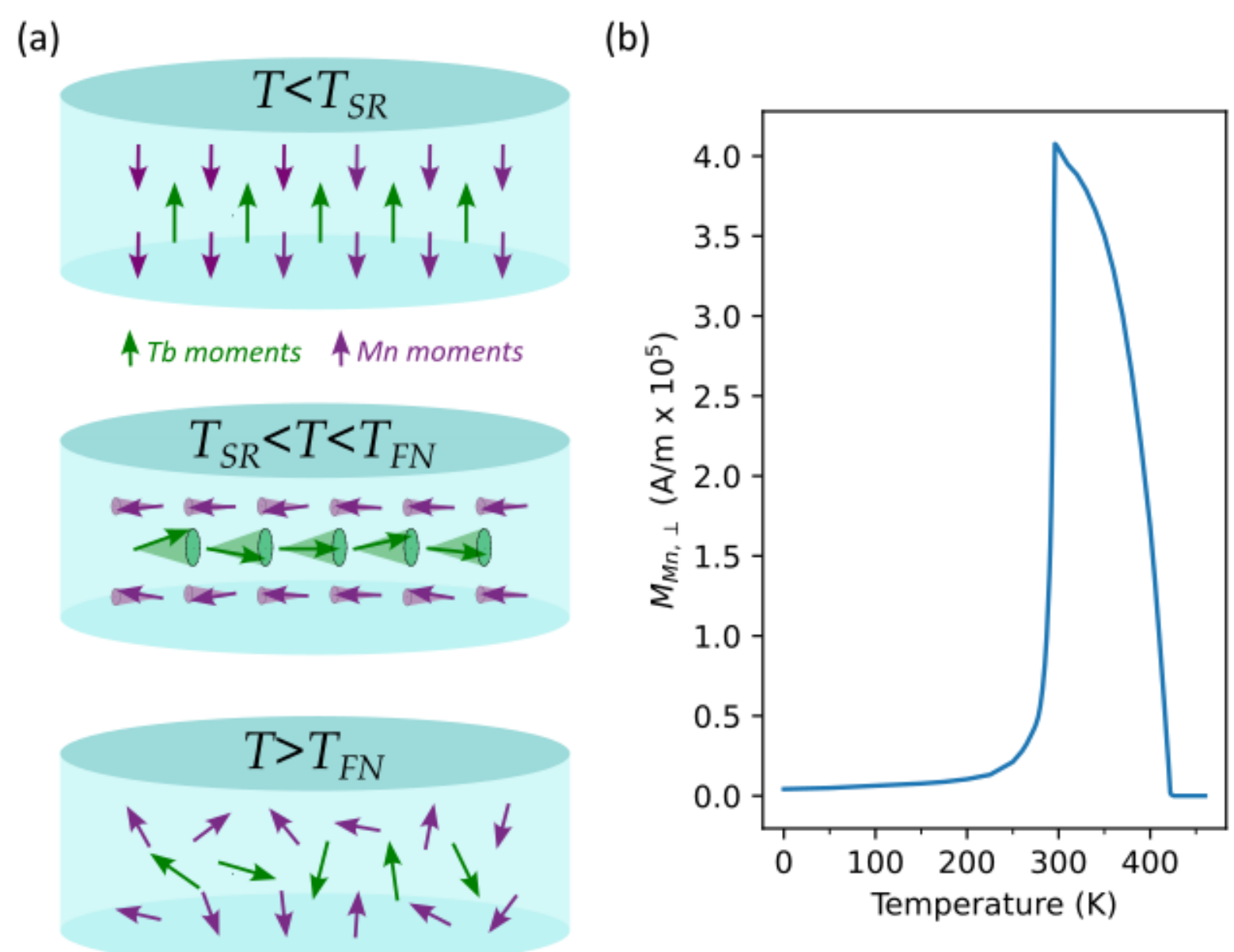

**Figure 1. Spin reorientation in $TbMn_6Sn_6$ (Tb166) as a function of temperature, under equilibrium conditions.** (a) Schematic of temperature-dependent spin states. There is an out-of-plane ferrimagnetic alignment between Mn and Tb moments below $T_{\mathrm{SR}}$. Above $T_{\mathrm{SR}}$, but below $T_{\mathrm{FN}}$, there is an in-plane ferrimagnetic alignment of Mn and Tb moments. As the temperature approaches $T_{\mathrm{FN}}$, spin-disorder increases due to thermal fluctuations, and this is represented with cones showing canted spins. The Tb disorder is depicted as being larger because the Tb moment drops off more rapidly than Mn as temperature is increased. Above $T_{\mathrm{FN}}$, the spin orientation is fully randomized. (b) The calculated in-plane projection of the Mn moment vs. temperature with an in-plane applied field of +185 mT (to match the experimental conditions). This subfigure was produced by combining anisotropy data from Jones *et al.* [5] and magnetic moment data from El-Idrissi *et al.* [10] – more details are provided in Section III. Modeling. The in-plane Mn moment is maximized just above the spin reorientation temperature. The signal is small or zero below $T_{\mathrm{SR}}$ and above $T_{\mathrm{FN}}$.

Technologically, Tb166 may be particularly promising for spintronic applications as the spin reorientation temperature is very near room temperature. Thus, it may be a candidate for a low power spintronic device. $TbMn_6Sn_6$ has also recently attracted substantial interest because the broader *R*166 family offers additional tunability through mixed orbital moments, anisotropies, and exchange couplings [16] as well as topological phenomena [9,17–20]. However, all previous studies of spin reorientation in Tb166 and related *R*166 compounds used static control [11,15,16,18,19,21–25], where for example, the sample is slowly heated or cooled across the spin reorientation transition. In this case, the new spin orientation is then inferred from equilibrium susceptibility or transport measurements.

Ultrashort laser pulses can rapidly excite spin dynamics on sub-picosecond timescales, [26] including spin transfer [27,28], ultrafast demagnetization [4], magnetic structure modification [4], and laser-induced spin reorientation [29,30]. When combined with ultrafast extreme UV (EUV) probes based on tabletop high harmonic generation or facility-scale soft X-ray sources, element-specific spin dynamics in magnetic multilayers and alloys can be uniquely tracked [31–34]. Past work has shown that transition metals typically demagnetize within a few hundred femtoseconds [26,35], whereas rare-earth lattices respond more slowly on picoseconds to tens of picoseconds timescales [35,36]. In rare-earth transition metals alloys, however, strong inter-sublattice exchange enables angular-momentum transfer between the 3d and 4f/5d states, producing demagnetization times comparable to those of pure TMs [37–39]. This is most pronounced near the magnetic compensation temperature, where the rare-earth and transition metal moments are equal and opposite [40]. $TbMn_6Sn_6$ lacks a compensation point because the magnetization of the Mn sublattice exceeds the Tb magnetization due to the 6:1 Mn:Tb ratio, despite Tb carrying a larger individual moment [5]. The strong Mn–Tb exchange and large anisotropy energies nonetheless suggest that demagnetization dynamics in this material could be significantly faster than in pure rare-earth materials, motivating direct time-resolved investigation. Element-specific probing at the Mn EUV *M*-edge is therefore particularly valuable for understanding spin dynamics in $TbMn_6Sn_6$.

Past research has shown that ultrafast laser excitation can drive spin reorientation in a variety of magnetic systems. Seminal work on the RE orthoferrite $TmFeO_3$ [29,41,42] and related insulating compounds [43–47] demonstrated that femtosecond pulses can trigger inertial spin switching and coherent magnon generation, typically detected via THz emission [29,48–50]. In metallic systems, laser-induced reorientation has primarily been explored in simple ferromagnetic films [51], ferromagnetic/antiferromagnetic bilayers [52], and organic molecular interfaces [53]. More recently, ultrafast optical control in spin reorientation materials has extended to the generation of topological spin textures such as skyrmions [54]. Despite their breadth, ultrafast spin reorientation studies have been primarily restricted to either insulating oxides or engineered heterostructures. Direct measurements of the reorientation timescale in a single-crystal metallic ferrimagnet, such as in Tb166, where the transition is driven by the intrinsic, temperature-dependent competition between two exchange-coupled magnetic sublattices, remains largely unexplored.

In this study, we directly measure the intrinsic timescale of the spin reorientation transition in $TbMn_6Sn_6$. We use a femtosecond laser to impulsively drive the spin system across the reorientation transition, and probe the resulting spin dynamics using time-resolved element-specific EUV magneto-optical spectroscopy [31–33]. By detecting weak TMOKE signals at the Mn *M*-edge, we determine that spin reorientation occurs over timescales between 6 and 18 ps, depending on the laser excitation fluence. This relatively fast response reflects the large magnetocrystalline anisotropy energies that govern the transition in Tb166, consistent with rare-earth–transition-metal ferrimagnets. A Landau–Lifshitz model incorporating temperature-dependent anisotropy fields reproduces these timescales and furthermore, predicts the possibility of 180° reorientations of the out-of-plane moment over a range of laser excitation fluences. This could facilitate optically controlled magnetization switching between very stable ground states, which could have useful applications in spintronics or data storage. These direct measurements of the timescale of spin reorientation in $TbMn_6Sn_6$ can serve as a benchmark to tune the response in similar compounds, to achieve even faster spin reorientation timescales.

In addition to measuring the timescale of spin reorientation, we also investigated laser-induced ultrafast demagnetization of $TbMn_6Sn_6$, that occurs at higher laser pump fluences. We find that demagnetization proceeds on a timescale of ~1 ps – an order of magnitude faster than spin reorientation. The clear separation of timescales and fluence dependence demonstrates that demagnetization and reorientation are distinct processes, with femtosecond laser excitation providing selective access to either regime.

## II. EXPERIMENTAL RESULTS

Time-resolved magnetization dynamics were measured on a Tb166 single crystal using the EUV transverse magneto-optical Kerr effect (TMOKE) at the Mn *M*-edge at the X-MATTER beamline [55]. The sample was excited by a 40–55 fs, ~800 nm laser pump pulse, and then probed using ~15 fs EUV pulses generated via high harmonic generation (HHG). The TMOKE magnetic asymmetry, $A = (I_+ - I_-)/(I_+ + I_-)$, was extracted from dynamic reflectivity measurements ($I_+$, $I_-$) performed while applying opposite in-plane magnetic fields (±185 mT). No active heating or cooling was applied to the sample. However, the sample was heated by thermal radiation from the electromagnet (situated behind the sample) as well as the pump laser (incident on the front face of the sample). See further details in the Supplemental Materials (SM), section S1.

In our TMOKE geometry, the asymmetry signal is sensitive to the in-plane component of the Mn magnetization, and the temperature dependence of this quantity is depicted in Figure 1(b). Therefore, any transient variation in the asymmetry reflects not only a possible change in the magnitude of the Mn moment but also any dynamic reorientation of the spin vector around the in-plane direction. As seen in Fig. 1(b), if the measured asymmetry signal has a value of zero or near-zero, this indicates that the sample temperature is either in the low temperature regime (below $T_{\mathrm{SR}}$) where no in-plane component exists, or the high temperature regime (above $T_{\mathrm{FN}}$) where no magnetic moment exists. In the first instance, the total moment is still large but points out-of-plane, and in the second instance, the total moment is zero.

We measured the time-dependent magnetic asymmetry signal for a range of pump fluences, using a high harmonic probing photon energy of 46.2 ± 0.2eV, which is close to resonance with the Mn *M*-edge at 47.2 eV [56]. Figure 2(a) plots example measurements of the spin reorientation dynamics in $TbMn_6Sn_6$ following femtosecond laser excitation at low pump fluences. The dynamics plotted in Fig. 2 can be separated into three regions. First, before laser excitation (time zero), the sample temperature is below the transition temperature $T_{\mathrm{SR}}$, so that the magnetization is predominantly out-of-plane, and the measured in-plane EUV MOKE signal is nearly zero. This is labelled step 1 in the schematic shown at the top of Fig. 2(a). In step 2, after laser excitation, Tb166 passes through the reorientation transition, and the TMOKE signal increases as spins rotate in-plane. The amplitude of the TMOKE response scales with pump fluence: higher fluences results in a more complete reorientation, corresponding to a larger fraction of the sample magnetized in-plane. In step 3, on timescales of ~30–200 ps and beyond, the deposited heat dissipates into the sample and mounting holder. As the sample cools below $T_{SR}$, the TMOKE signal decreases as the spins gradually relax back to an out-of-plane orientation.

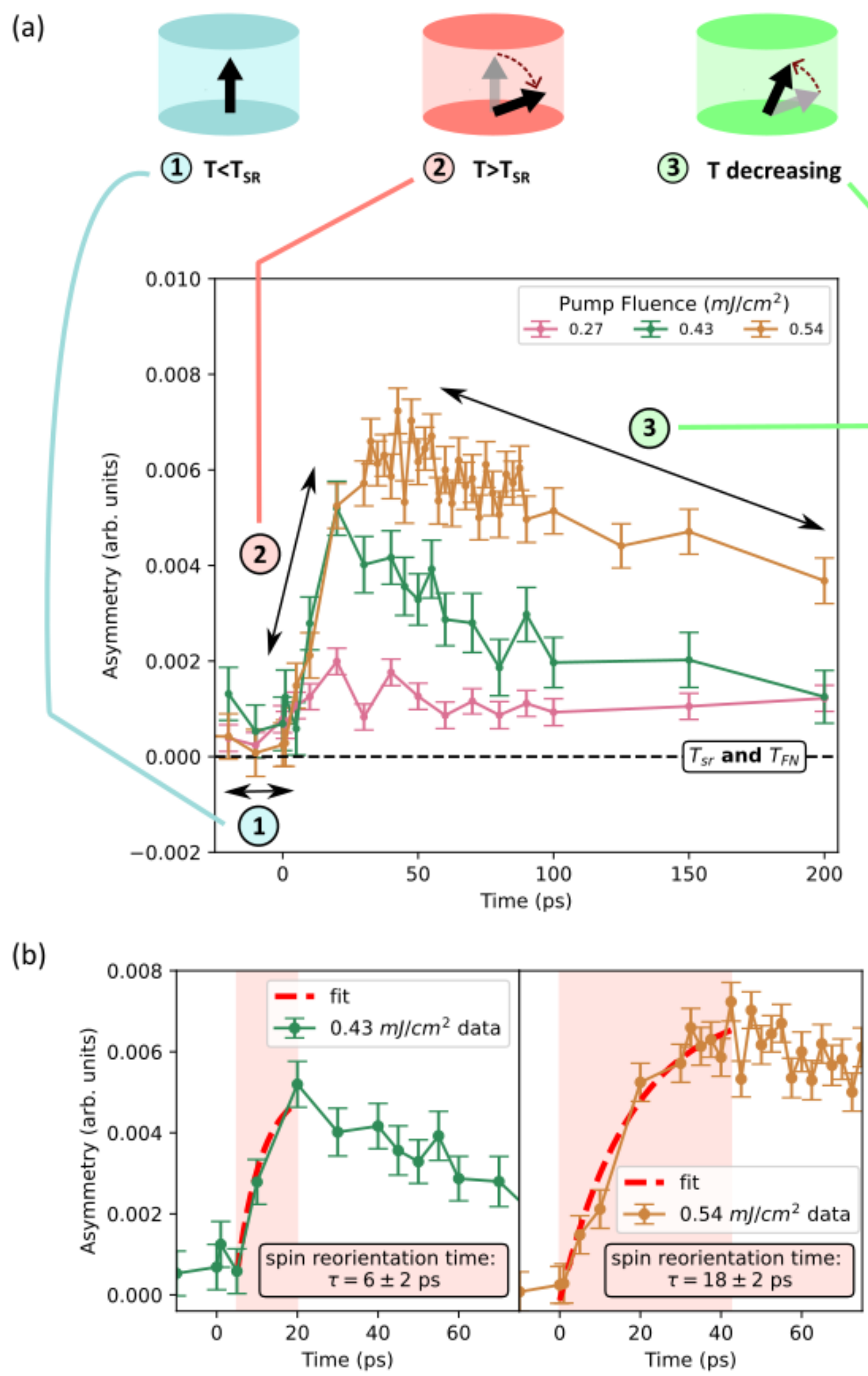

**Figure 2. Spin reorientation dynamics in $TbMn_6Sn_6$.** (a) Measured dynamics of the in-plane Mn asymmetry following low-fluence laser excitation. The horizontal black dotted line indicates zero in-plane Mn moment, which can occur either when the temperature is below the spin-reorientation temperature ($T_{SR}$, magnetization fully out-of-plane) or above the ferrimagnetic Néel temperature ($T_{FN}$, long-range order vanishes). The low-fluence dynamics proceed in three stages as illustrated in the upper cartoon: (1) prior to pump excitation, the sample temperature is below $T_{SR}$ and the signal is approximately zero; (2) upon laser excitation, the sample is transiently heated above $T_{SR}$,

and the magnetization begins to rotate in-plane, causing the signal to increase; (3) as the sample cools, the in-plane magnetization gradually decreases. (b) Fits to the data yield spin reorientation times of 6 ± 2 ps and 18 ± 2 ps based on Equation (1), demonstrating that the reorientation occurs on few to tens-of-picoseconds timescales.

In Fig. 2(b), we fit the low fluence data from Fig. 2(a), with a bounded exponential of the form:

$$A(t) = A_{\max}\left[1 - e^{\frac{-(t-t_0)}{\tau}}\right] \quad (1)$$

where $A(t)$ is the magnetic asymmetry signal, $A_{\max}$ is the maximum measured signal amplitude for each fluence as measured in Fig. 1, $t$ is time, $t_0$ is an offset to account for any differences between the time zero as defined in the figure and the time zero determined by the fit. $\tau$ is the time-constant of the exponential term, which reflects the spin reorientation time. For fitting, data was truncated to include only the exponential rise (from the lowest to the highest point) to isolate the spin reorientation transition from the subsequent cooling stage. The extracted time constants were 6 ± 2 ps at 0.43 mJ/cm² and 18 ± 2 ps at 0.54 mJ/cm². The very low 0.27 mJ/cm² data was not fit due to poor signal-to-noise ratio. The longer fitted timescales at higher fluences can be explained by temperature-dependent changes in the anisotropy field, magnetic moment and the resultant torque. As the temperature approaches $T_{\mathrm{FN}}$, the torque towards the in-plane direction is reduced since magnetic disorder reduces the strength of the anisotropy field. This trend is also replicated in the modelling section.

We attribute the ultrafast (<20 ps) reorientation time to the strong magnetic anisotropy in $TbMn_6Sn_6$ (Jones *et al.* [5]), which defines the rapid rotation of the spin orientation. Finally, we note that in both Figs. 2 and 3, a fluence-dependent constant scaling factor between 0.8-1.8 was applied to each curve to account for the experimentally measured fluence-dependent sample degradation. More details of this measurement are provided in SM, section S2.

At still higher laser pump fluences, the sample exhibits markedly different dynamics, as shown in Fig. 3. Panel (a) depicts the time-resolved TMOKE signal for laser pump fluences of 1.1, 2.1, and 3.6 mJ/cm$^2$. Here, the initial sample temperature before laser excitation is higher than in the low-fluence case (Fig. 2) due to stronger laser pump heating and the lack of active cooling. Given the 5 kHz laser repetition rate, the sample cannot fully cool below $T_{\mathrm{SR}}$ between laser excitation pulses. Consequently, a larger fraction of Mn moments is initially in-plane, producing a non-zero in-plane asymmetry before the laser pump pulse is incident (this is depicted in step 1 of Fig. 3). After laser excitation, the sample subsequently demagnetizes (step 2), as the electron–spin temperature exceeds $T_{\mathrm{FN}}$. At still later times, the Tb166 spin system cools primarily due to heat transport, causing it to remagnetize in-plane (step 3) as the sample temperature drops below $T_{\mathrm{FN}}$, while remaining above $T_{\mathrm{SR}}$. The recovery time increases with laser fluence, with 3.6 mJ/cm² recovering the slowest. As the sample cools further toward $T_{\mathrm{SR}}$, out-of-plane order returns and the signal decreases (step 4). This occurs on timescales of 200 ps for laser excitation fluences of 1.1 mJ/cm². Even longer recovery times are expected for higher laser fluences (timescales not captured in Fig. 3). In this high-fluence regime, the spin dynamics are dominated by the in-plane remagnetization and sample cooling processes (steps 3 & 4 in Fig. 3(a)). Thus, the recovery timescales reflect the rate of heat dissipation from the laser-excited region.

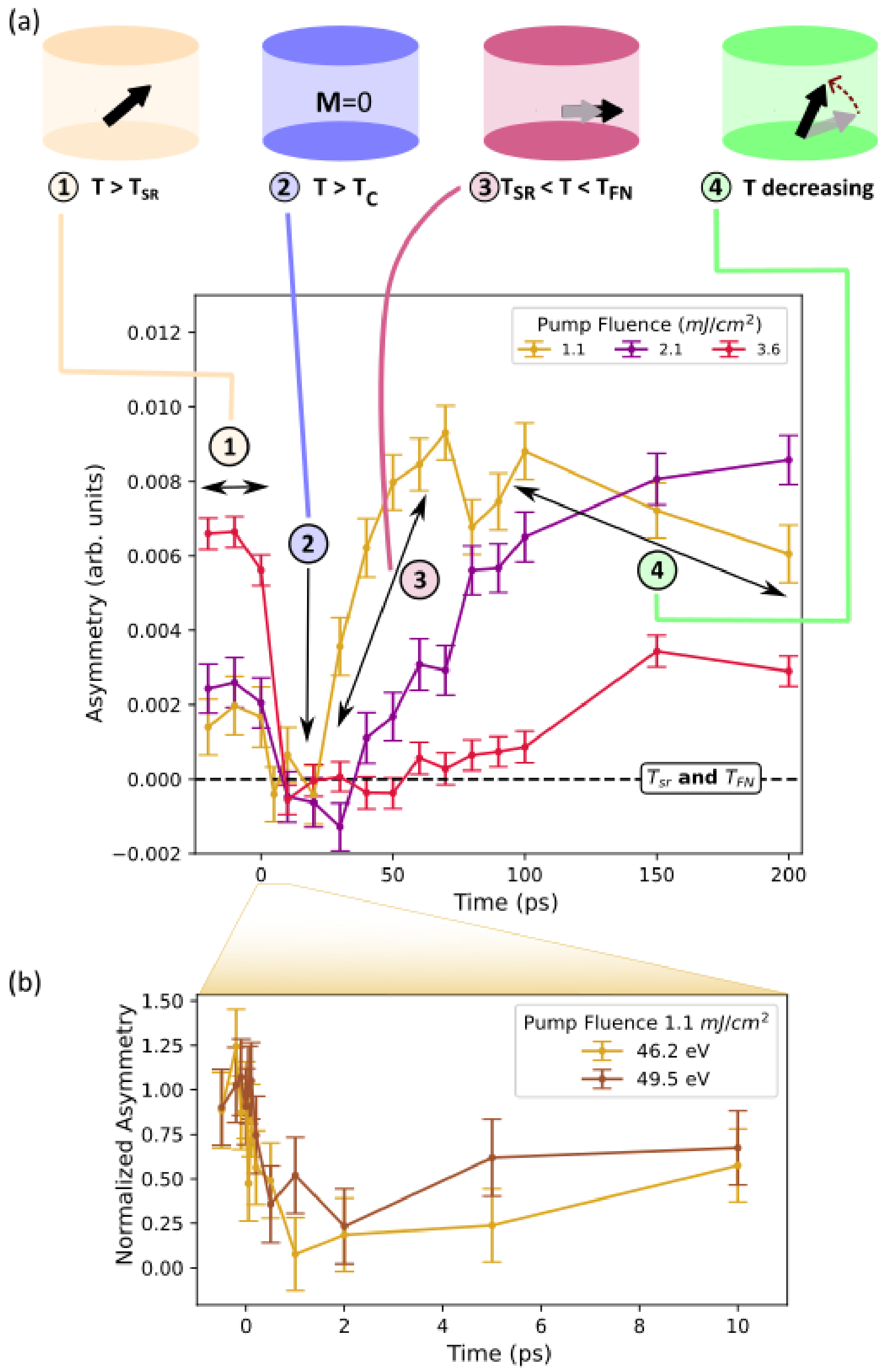

**Figure 3. High fluence spin reorientation and demagnetization dynamics in $TbMn_6Sn_6$.** (a) The measured in-plane Mn asymmetry following high-fluence laser excitation. Four stages of dynamics are depicted in the upper cartoon. (1) The sample temperature is slightly above $T_{SR}$ and the magnetization is initially partially orientated in-plane. (2) Following laser excitation, the sample is transiently heated above $T_{FN}$ and fully demagnetizes. (3) The sample cools below $T_{FN}$ and the magnetization recovers in an in-plane orientation. (4) This step is only present within the measurement timeframe on the 1.1 mJ/cm$^2$ data. Further cooling causes the magnetization to in-plane magnetization to decrease. (b) Demagnetization of Tb166 for a pump fluence of 1.1 mJ/cm$^2$ on shorter time scales, measured at two different harmonic probing energies. Demagnetization is an order of magnitude faster than the spin reorientation timescale. Dynamic asymmetries from -0.5 to 10 ps are measured with harmonic probing energies centered at 46.2 eV ± 0.2 eV and 49.5 eV ± 0.2 eV.

Panel (b) of Fig. 3 zooms in on the response on sub-10 ps timescales, comparing data taken with probe harmonics at the resonant *M*-edge, 46.2 eV, as well as at the off-resonant energy, 49.5 eV, with a pump fluence of 1.1 mJ/cm². The pre-time-zero signal was normalized to unity. Both measurements exhibit similar behavior, demonstrating that the dynamics are not particularly sensitive to the specific Mn *M*-edge probing energy, where the ultrafast demagnetization proceeds on a timescale of approximately 1 ps. As expected, this is significantly faster than the demagnetization time measured for pure Tb (8 ps) [36] due to pathways whereby angular momentum can transfer between 3d and 4f orbitals [40]. We note that this is longer than that in typical Mn-based metallic systems (∼100–300 fs) [57], where the dynamics are dominated by itinerant 3d electrons. This places $TbMn_6Sn_6$ in a regime, similar to rare-earth–transition-metal compounds (e.g., GdFeCo [58], TbFeCo [59]), where distinct demagnetization timescales arise from the interplay between transition-metal 3d and localized 4f moments. The demagnetization timescale is an order of magnitude faster than the spin reorientation dynamics discussed in Fig. 2. These measurements clearly distinguish the ultrafast demagnetization channel from the slower tens-of-picoseconds spin reorientation process.

## III. MODELING

To further understand laser-induced spin reorientation, we employed a simplified thermal-magnetic model based on the Landau-Lifshitz equation. The time-dependent effective field, $\mathbf{H_{eff}}(t)$ , which will act on the sample's magnetization unit vector, $\mathbf{m}(t)$, is described by:

$$\mathbf{H_{eff}}(t) = \mathbf{H_{appl}} + \mathbf{H_{demag}}(t) + \mathbf{H_{anis}}(t) \tag{2}$$

where $\mathbf{H_{appl}}$ is the constant field applied in the +*x* direction, with a value of 185 mT. Calculations of the demagnetizing field, $\mathbf{H_{demag}}(t)$, are given in more details in the Supplemental Materials, section S4, but have very little influence on the dynamics. The anisotropy field, $\mathbf{H_{anis}}(t)$ acts along the *z*-axis and is described by the following equation, with more details given in the Supplementary Information, S4:

$$\mathbf{H_{anis}}(t) = m_z(t)\left[\frac{\mathrm{MAE_{Tb}}\left(T_{\mathrm{spin}}(t)\right) + 6\mathrm{MAE_{Mn}}\left(T_{\mathrm{spin}}(t)\right)}{\mu_0 M_s\left(T_{\mathrm{spin}}(t)\right)}\right]\hat{\mathbf{z}} \tag{3}$$

where $m_z(t)$ is the *z*-projection of $\mathbf{m}(\mathrm{t})$ and $M_\mathrm{s}(T_{\mathrm{spin}}(t))$ is the saturation magnetization determined from neutron scattering [10] as a function of the instantaneous spin-bath temperature $T_{\mathrm{spin}}(t)$. The $M_\mathrm{s}(T_{\mathrm{spin}}(t))$ term accounts for the partial demagnetization of the system as the spin temperature is increased. The temperature-dependent magnetic anisotropy energies of the Mn and Tb sublattices, $\mathrm{MAE_{Mn}}$ and $\mathrm{MAE_{Tb}}$, follow Jones *et al.* (2024) and include the exchange field from the Mn-Tb interaction [5]. The usual factor of two converting anisotropy energy to field is omitted for consistency with Ref. 5, and a factor of six accounts for the six Mn atoms per unit cell.

The laser excitation is assumed to induce a transient rise in the spin temperature $T_{\mathrm{spin}}$, which modifies the temperature-dependent magnetic anisotropy and Mn-Tb exchange interactions. The temporal evolution of the spin system temperature, $T_{\mathrm{spin}}(t)$, is modeled phenomenologically as a sum of three exponential functions:

$$T_{\mathrm{spin}}(t) = T_{\mathrm{initial}} + T_{\mathrm{excite}}\left(-e^{-t/\tau_{\mathrm{rise}}} + Ae^{-t/\tau_{\mathrm{decay1}}} + (1-A)e^{-t/\tau_{\mathrm{decay2}}}\right) \tag{4}$$

where $T_{\text{initial}}$ is the initial sample temperature prior to the pump laser excitation (i.e. room temperature). The exponential term (amplitude $T_{\text{excite}}$, timescale $\tau_{\text{rise}} \approx 1$ ps), describes the rapid heating of the spin subsystem from the laser-excited electrons. The timescale of $\tau_{\text{rise}} \approx 1$ ps is in keeping with the demagnetization timescale measured in Fig. 3(b) where the spin bath is shown to be capable of heating above $T_{\text{FN}}$ on a 1 ps timescale. Within this phenomenological framework, $\tau_{\text{rise}}$ should be understood as an effective parameter describing the temporal evolution of the spin temperature, rather than a direct measure of a specific microscopic processes. Accordingly, this model captures the energy flow into the spin subsystem but does not explicitly describe angular momentum transfer processes. The temperature model in Eq. 4 describes an initial increase in spin temperature on short timescales which decays back to $T_{\text{initial}}$ on longer timescales. The sample cools with two timescales $\tau_{\text{decay1}}$ and $\tau_{\text{decay2}}$. The amplitude of these two decay functions are $A$ and $(1 - A)$ where $0 < A < 1$. The second exponential term, described by the timescale $\tau_{\text{decay1}}$ accounts for spin–lattice thermalization through spin–phonon coupling on a timescale of $\approx$ 3 ps. The final exponential described by $\tau_{\text{decay2}} \approx 200$ ps corresponds to cooling through heat diffusion into the substrate and environment, where the value of 200 ps serves as an approximate upper bound consistent with the observed decay duration of the TMOKE signal. The timescales for $\tau_{\text{rise}}$ and $\tau_{\text{decay1}}$ are representative estimates based on typical literature values [60,61]. Nevertheless, this simplified model reproduces the characteristic heating and relaxation behavior observed experimentally.

The model predicts a spin reorientation near 300 K in zero-field and at slightly lower temperature (≈ 295 K) under an in-plane field of 185 mT. This is close to the literature value for spin reorientation of 309 K [5], although the model is slightly lower than the literature value. It is also lower than our experimentally inferred spin-reorientation temperature of 314 K based on the pre-time-zero asymmetry (Supplemental Materials, Fig. S1), where this measured temperature may differ slightly from the actual sample temperature due to the placement of the temperature probe, as discussed in Supplemental Materials, section S1. We therefore initialized the simulation at $T_{\text{initial}}$ = 270 K, allowing the heating pulse to drive the system, $T_{\text{spin}}(t)$, through the transition during the modeled simulation. This small discrepancy in $T_{\text{SR}}$ does not affect the fundamental dynamics, the model successfully reproduces the main experimental trends and provides a consistent framework for interpreting the heating-driven reorientation (Fig. 4).

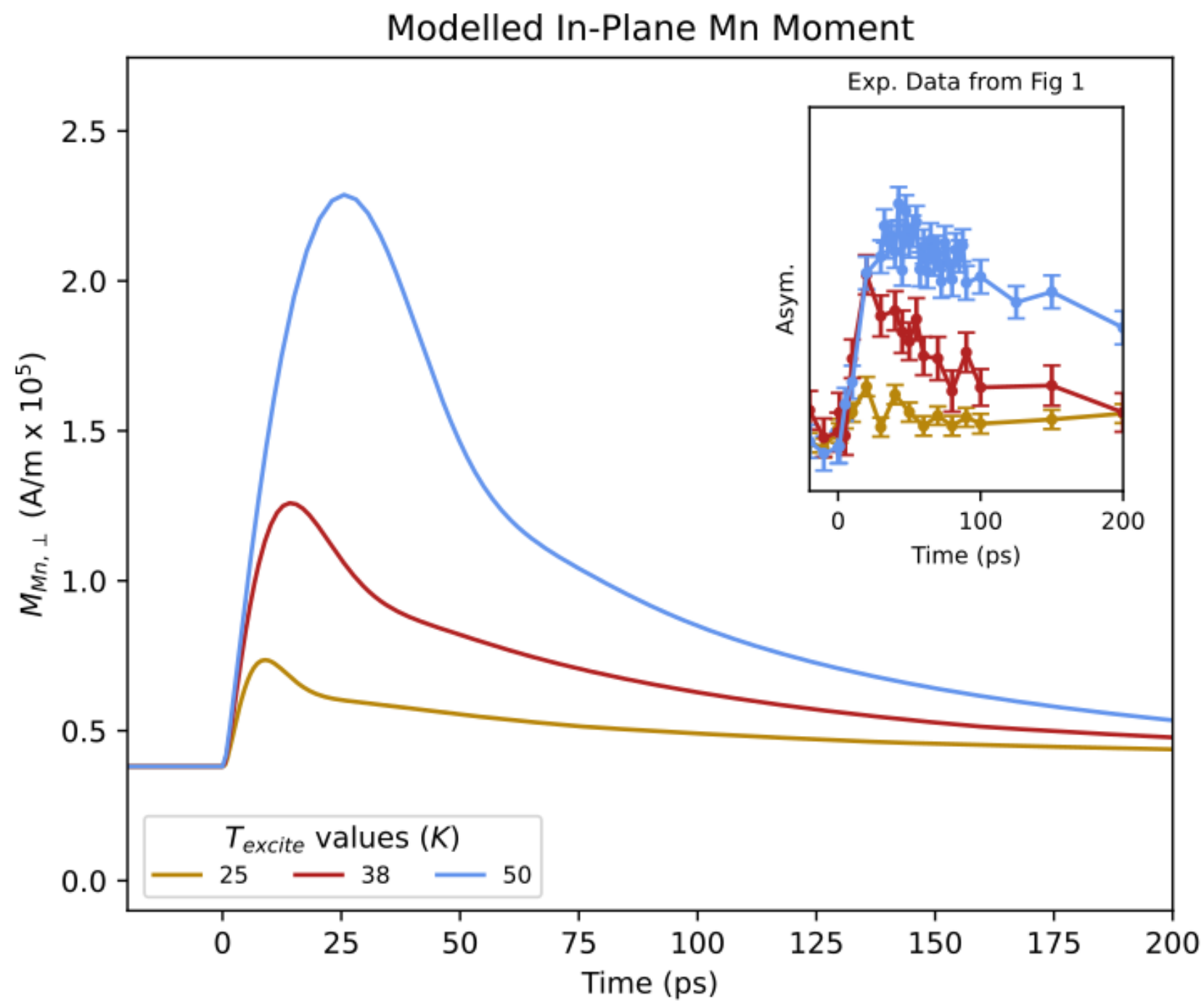

**Figure 4. A dynamic model of spin orientation in Tb166 in a low pump fluence range.** The modeled signals explore the effect of varying the initial heating amplitude $T_{\text{excite}}$, based on the temperature profile described by Eq. (4). The model successfully reproduces the overall trend of the experimental curves (shown in the inset).

The temperature-dependent anisotropy field is then incorporated into the Landau–Lifshitz equation:

$$\frac{d\mathbf{m}}{dt} = -\gamma\left(\mathbf{m} \times \left(\mathbf{H_{demag}}(t) + \begin{bmatrix} H_{\text{appl}} \\ 0 \\ H_{\text{anis}}(t) \end{bmatrix}\right)\right) - \frac{\alpha\gamma}{m}\left(\mathbf{m} \times \mathbf{m} \times \left(\mathbf{H_{demag}}(t) + \begin{bmatrix} H_{\text{appl}} \\ 0 \\ H_{\text{anis}}(t) \end{bmatrix}\right)\right) \quad (5)$$

where $\gamma$ is the gyromagnetic ratio and $\alpha$ is the damping factor. The equation is solved numerically using MATLAB's ODE45 solver. For comparison with the EUV-TMOKE data, the simulated in-plane magnetization component $m_x(t)$ is multiplied by the temperature-dependent Mn magnetization. The amplitude of the spin-temperature rise ($T_{\text{excite}}$ in Eq. 4) is equivalent to a simulated fluence as this parameter controls the heating of the sample. Unless otherwise stated, the parameters are $A = 1/2$, $\tau_{\text{rise}} = 1$ ps, $\tau_{\text{decay1}} = 3$ ps, $\tau_{\text{decay2}} = 200$ ps, and $\alpha = 1$. Modeling with different timescales and amplitudes (see the Supplemental Materials, section S5) shows that the reorientation behavior is relatively insensitive to the exact parameters used.

The simulations reproduce the main experimental features—a monotonic rise and gradual decay of $m_x(t)$ of $\approx 200$ ps—consistent with heating-dominated anisotropy softening. A critical damping factor of $\alpha = 1$ was adopted to suppress the large oscillations that appear for smaller values ($\alpha = 0.01$–$0.1$) shown in the Supplemental Materials (Fig. S2(h)). This is consistent with the absence of coherent precession in the experimental data (Fig. 3). The absence of clear precession in the experimental data suggests spatially nonuniform excitation and sample inhomogeneities that wash out any coherent oscillations in the probed volume.

Fitting the simulated curves in Fig. 3 using equation (1) yields characteristic rise times of 3.8 ± 0.1 ps, 5.6 ± 0.1 ps, and 9.8 ± 0.2 ps for heating amplitudes $T_{\mathrm{excite}}$ = 25, 38, and 50 K, respectively. This is in good agreement with the experimental timescales of 6 ± 2 ps and 18 ± 2 ps. With an $\alpha$ value of 0.1, which allows for strong precessions, as depicted in the Supplemental Materials Fig. S2(h), the rise times obtains were 9.0 ± 0.4 ps, 15.0 ± 0.5 ps, and 28.3 ± 0.7 ps. In this $\alpha$ = 0.1 model, the fitted reorientation timescales for the three values chosen for $T_{\mathrm{excite}}$ increased by a factor of 2-3 as compared to $\alpha$ = 1. However, the values obtained of 9-28 ps are still within good agreement of the experimental values of 6-18 ps. While the exact value of $\alpha$ in unknown in this material and is expected to vary strongly with temperature, the range of timescales measured in the experiment are generally reproduced by the model regardless of the chosen $\alpha$ value.

To verify that these modeled timescales reflect an intrinsic spin-reorientation response rather than details of the empirical temperature model, we also tested a step-function heating profile, in which the spin temperature was instantaneously increased to its maximum value without gradual rise or decay. The before and after temperatures in the step-function were chosen to match the initial and peak temperatures from Fig 4. The step-function model is plotted in SM Fig S2. The resulting fits of 4.8 ± 0.1 ps, 15.5 ± 0.1 ps, and 21.2 ± 0.1 ps overlap with the experimental values (6 - 18 ps). This indicates that the extracted timescales reflect intrinsic spin-reorientation dynamics. The characteristic picosecond reorientation can be rationalized by the energy–time relation [62], $t \approx h/E$. For an anisotropy energy on the order of a few meV—an expected range for $TbMn_6Sn_6$ [5] the corresponding period is several picoseconds ($\approx$ 8 ps for 2 meV), consistent with the measured and simulated timescales. Note that the exact value of the anisotropy energy varies strongly with temperature.

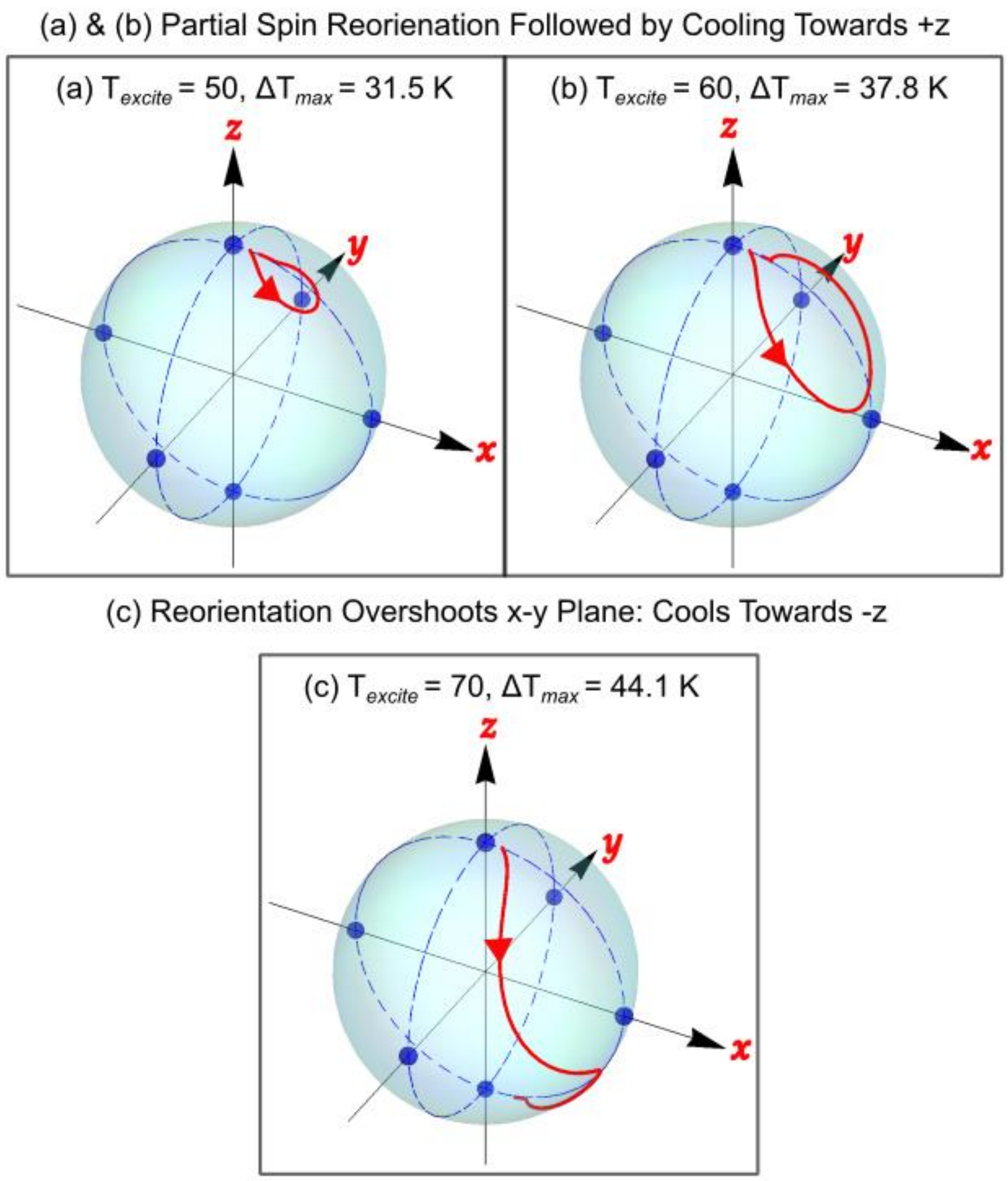

**Figure 5. 3-dimensional trajectories of the Tb166 magnetization vector obtained from Landau–Lifshitz simulations.** The red curves trace the path of the magnetization unit vector over the first 200 ps of the simulated dynamics. The sample plane lies in the *x*–*y* plane, while the *z*-axis corresponds to the out-of-plane direction. Results are shown for three different amplitudes of laser-induced temperature increase: (A) $T_{\text{excite}} = 50$, $\Delta T_{\max} = 31.5$ K; (B) $T_{\text{excite}} = 60$, $\Delta T_{\max} = 37.8$ K; and (C) $T_{\text{excite}} =70$, $\Delta T_{\max} = 44.1$ K. Increasing the excitation amplitude $T_{\text{excite}}$ drives progressively larger angular excursions of the magnetization, indicating a transition from partial to nearly complete reorientation as shown in subfigures (a) and (b). Subfigure (c) depicts a 180° spin rotation that can occur if the spin reorientation overshoots the in-plane (*x*-*y*) direction and ultimately cools towards the -*z* direction.

Figure 5 illustrates the three-dimensional nature of the simulated spin reorientation as the sample heating increases. The initial temperature is 270 K, meaning that the spin is orientated out-of-plane, i.e. on the *z*-axis. The applied field of 185 mT in the +*x* direction slightly cants the moment toward the *x*-axis. At moderate fluences ($T_{\text{excite}}$ = 50–60 K), the magnetization vector traces a small circular trajectory before cooling back toward its initial orientation. At higher simulated fluences ($T_{\text{excite}} \geq 70$ K, Panel c), the spin trajectory overshoots the *x*-*y* plane due to the large reorientation torque and continues into the –*z* hemisphere. As the system cools back below $T_{\text{SR}}$ while the *z*-component is negative, the magnetization relaxes into the energetically equivalent –*z* state—resulting in a 180° reversal of the ground-state direction. This bidirectional switching between ±*z* room temperature ground states suggests that $TbMn_6Sn_6$ can, in principle, be toggled between two stable magnetic configurations with modest laser fluences near room temperature. The final orientation, however, is expected to depend sensitively on the damping and the detailed thermal profile, and the EUV-TMOKE geometry used here cannot distinguish between the ±$z$ alignments. For example, with a value of $\alpha$ = 0.1 (instead of $\alpha$ = 1), the behavior in Fig 5(a) 50K, and (c) 70 K, are replicated with additional oscillations that stabilize within ~200 ps. However, for the 60 K case (Fig 5(b)), the oscillations at $\alpha$ = 0.1 cause the vector to overshoot the in-plane direction. In this case, the final equilibrium spin direction will match that of Fig 5(c), i.e. in the -z direction. Therefore, with alpha values of <1, the spin direction toggling behavior can be induced with smaller temperature changes.

## IV. DISCUSSION

These results represent the first direct investigation of the intrinsic timescale of the spin reorientation transition in the Kagome ferrimagnet $TbMn_6Sn_6$, in this case driven by femtosecond-laser excitation rather than static heating. From the time-resolved EUV-TMOKE measurements, we extract a characteristic reorientation timescale of 6-18 ps, depending on the laser excitation fluence. Furthermore, we can selectively excite and probe two key ultrafast processes: a tens-of-picoseconds spin reorientation of the Mn moments that occurs at low excitation fluences, and demagnetization of the Mn moment on a faster picosecond timescale, that occurs at higher laser excitation fluences.

The picosecond demagnetization timescale is consistent with rapid angular-momentum transfer between the Mn 3d and Tb 4f sublattices and the electronic bath – as expected for ferrimagnetic systems – even though $TbMn_6Sn_6$ lacks a compensation point. In contrast, the slower (10–20 ps) spin-reorientation response reflects laser-induced changes in the competing Mn and Tb anisotropy fields: transient heating above $T_{\mathrm{SR}}$ modifies the balance between the in-plane Mn and out-of-plane Tb anisotropy preferences. This drives a collective rotation toward the new equilibrium orientation. This timescale is consistent with reorientation governed by large anisotropy energies on the meV scale.

Unlike rare-earth orthoferrites [29], where spin orientation is typically associated with the coherent optical torques such as the inverse Faraday effect (IFE), the dynamics in $TbMn_6Sn_6$ are dominated by thermally assisted anisotropy modulation rooted in the intrinsic competition between the magnetic sublattices. In this study, spin reorientation is achieved with linearly polarized excitation—under which helicity dependent IFE-like torques are absent—and the experimental results are well reproduced by a heating-driven model. This establishes a purely thermal pathway to control spin orientation using modest laser fluences.

We do not implement the model in the high fluence regime (demagnetizing) regime of Fig. 3, since the standard LL equation becomes less accurate near the ferrimagnetic Néel temperature, where the magnetization magnitude and damping vary strongly [63]. A microscopic Landau–Lifshitz–Bloch approach [64] would be required to describe this regime, and can be a topic for future work.

A Landau–Lifshitz model incorporating temperature-dependent anisotropy quantitatively reproduces both the experimental timescales (6–18 ps) for the spin orientation and their fluence dependence, verifying that the reorientation rate is set by the transient anisotropy strength. At higher excitations, the model predicts an overshoot of the magnetization across the plane and a possible 180° reversal of the out-of-plane component, implying that $TbMn_6Sn_6$ could, in principle, be switched between two equivalent magnetic ground states by moderate laser heating.

Finally, we note that the EUV-TMOKE geometry is insensitive to the sign of the out-of-plane moment, so the ±$z$ orientations cannot be distinguished. Nevertheless, the potential for reversible 180° reorientation highlights $TbMn_6Sn_6$ as a promising platform for thermally assisted optical control of magnetization. The large anisotropy barrier and relatively low $T_{SR}$ enable switching with small temperature excursions compared with conventional transition-metal alloys, providing a new avenue for ultrafast spin manipulation in Kagome magnets.

## V. CONCLUSIONS

In conclusion, by measuring element-specific, transverse magneto-optic Kerr-effect spectroscopy signals across the Mn *M*-edge, we directly probe the ultrafast spin reorientation and demagnetization dynamics in $TbMn_6Sn_6$ for the first time. Distinct dynamical behaviors are observed for low and high laser excitation fluences. For low laser fluences, spin reorientation proceeds on timescales of 6-18 ps, depending on the exact excitation fluence, and is well reproduced by a Landau–Lifshitz model incorporating a temperature-dependent magnetocrystalline anisotropy field. Finally, our model predicts a possibility of a 180° reorientation of the out-of-plane moment, that could facilitate optically controlled magnetization switching between very stable ground states, for applications in spintronics or data storage. These results establish a direct benchmark for the intrinsic timescale of spin reorientation in $TbMn_6Sn_6$ and provide a foundation for tuning relevant energy scales through chemical substitution, as well as for exploring potential spintronic and data storage applications.

## ACKNOWLEDEMENTS

The authors acknowledge Igor I. Mazin for insightful discussion.

## FUNDING

Department of Energy Office of Basic Energy Sciences X-Ray Scattering Program under Award No. DE-SC0002002 (H.C.K., M.M.M.). National Science Foundation (NSF) CAREER award DMR- 2343536 (N.J.G.)

## AUTHOR CONTRIBUTIONS

S.A.R., A.G., N.L., T.R.K., R.S., H.C.K. and M.M.M. conceived and conducted the experiments, and analyzed the experimental data. J.M.S. and H.N. advised on the theoretical calculations performed by S.A.R. The sample was grown and characterized by H.B. and N.J.G.. N.L. made major contributions to the conceptual framework and led the writing of the manuscript. All authors contributed to writing the manuscript.

## COMPETING INTERESTS

H.C.K. and M.M.M. have a financial interest in a laser company, KMLabs, that produces the laser used in this work. H.C.K. is partially employed by KMLabs. All other authors declare they have no competing interests.

Certain commercial equipment, instruments, or materials are identified in this paper in order to specify the experimental procedure adequately. Such identification is not intended to imply recommendation or endorsement by the National Institute of Standards and Technology, nor is it intended to imply that the materials or equipment identified are necessarily the best available for the purpose.

Supplementary Materials for

# Direct Measurement of the Timescale of Spin Reorientation in $TbMn_6Sn_6$

Sinéad A. Ryan *et al.*

*Corresponding author: Na.li-1@colorado.edu

**Contents:**

## Supplementary Text

### Section S1. Experimental Design

Measurements of time-resolved magnetization dynamics on $TbMn_6Sn_6$ single crystal were performed using the X-MATTER beamline [55] in Boulder. A Ti: Sapphire regenerative amplifier was used to drive high harmonic generation (HHG) in a neon gas medium. This produced a comb of harmonic probing energies in the extreme ultraviolet (EUV) range. Energies in the HHG comb were spectrally dispersed using a diffraction grating and the intensities were recorded on a CCD chip. The EUV light resonantly probed the in-plane Mn moment at the *M*-edge (47.2 eV) using the transverse magneto-optical Kerr effect (TMOKE). Sample dynamics were excited using pulses from the Ti: Sapphire laser with a 780 nm central wavelength, 40 fs duration pump laser and a 5 kHz repetition rate. The pump spot had $1/e^2$ radii of 316 μm by 535 μm. The probe spot had a $1/e^2$ radius of approximately 40 μm.

No active heating or cooling was applied to the sample. However, larger pump fluences resulted in higher baseline sample temperatures due to the undissipated heat from the pump laser.

The magnitude of the in-plane Mn magnetic moment was determined by calculating the TMOKE magnetic asymmetry, which was based on differential intensity measurements with applied fields of 185 mT ± 15 mT in-plane and a *p*-polarized incident beam at near-Brewster:

$$A(t) = \frac{I_+(t) - I_-(t)}{I_+(t) + I_-(t)} \quad (1)$$

Where $A(t)$ is the magnetic asymmetry signal, $I_+(t)$ is the time-dependent reflectivity of the sample with an applied field of +185 mT, and $I_-(t)$, is the time-dependent reflectivity of the sample with an applied field of -185 mT. More details of the beamline design and magnetic asymmetry measurement scheme appear in Johnsen *et al.* [55].

Single crystals of $TbMn_6Sn_6$ were grown by self-flux methods as described by Jones *et al*. [5]. The sample was mounted on a 0.5 mm thick sapphire plate using a conductive silver epoxy (EPO-TEK® H20S). No active heating or cooling systems were applied to the sample. Sample temperatures were monitored using an in-vacuum thermocouple wire mounted on the back of the sapphire sample plate.

A maximal in-plane magnetic field of 185 ± 15 mT was applied using an out-of-vacuum GMW 5201 projection field electromagnet. The in-plane applied field was maximized to encourage the reorientation transition. The electromagnet was water cooled but still reached temperatures exceeding 40°C during use. The sample was heated to above room temperature by thermal radiation from the electromagnet (situated behind the sample) and the pump laser (incident on the front of the sample). For each measurement, we allowed the sample to the thermalize for approximately 1 hr with the same electromagnet currents and laser pump fluences used in the experiment in order for the sample to reach a stable baseline temperature before data was taken. The electromagnet alone increased the sample temperature by approximately 18 K (to ~310 K). Additional pump-induced heating was fluence dependent: measured temperatures ranged from ~ 319 K at the highest fluence (3.6 mJ/cm²) to ~311 K at the lowest fluence (0.27 mJ/cm²) (Fig. S1). There was a small amount of ongoing thermalization that continued to increase the sample temperature throughout the scan after the 1 hr thermalization period. For this reason, we measured the temperature at the beginning (slightly lower temperature) and end of the scan (slightly higher temperature). These high and low temperature set the upper and lower bounds of the bars in Fig. S1. Heating from the EUV probe is negligible.

The pre-time-zero asymmetry is also plotted as a function of pump fluence in Fig. S1. When combined with the thermocouple measurements, this analysis provides an approximate estimate of the spin-reorientation temperature, which appears roughly 5 K higher than the value obtained via static Superconducting Quantum Interference Device (SQUID) magnetometer.

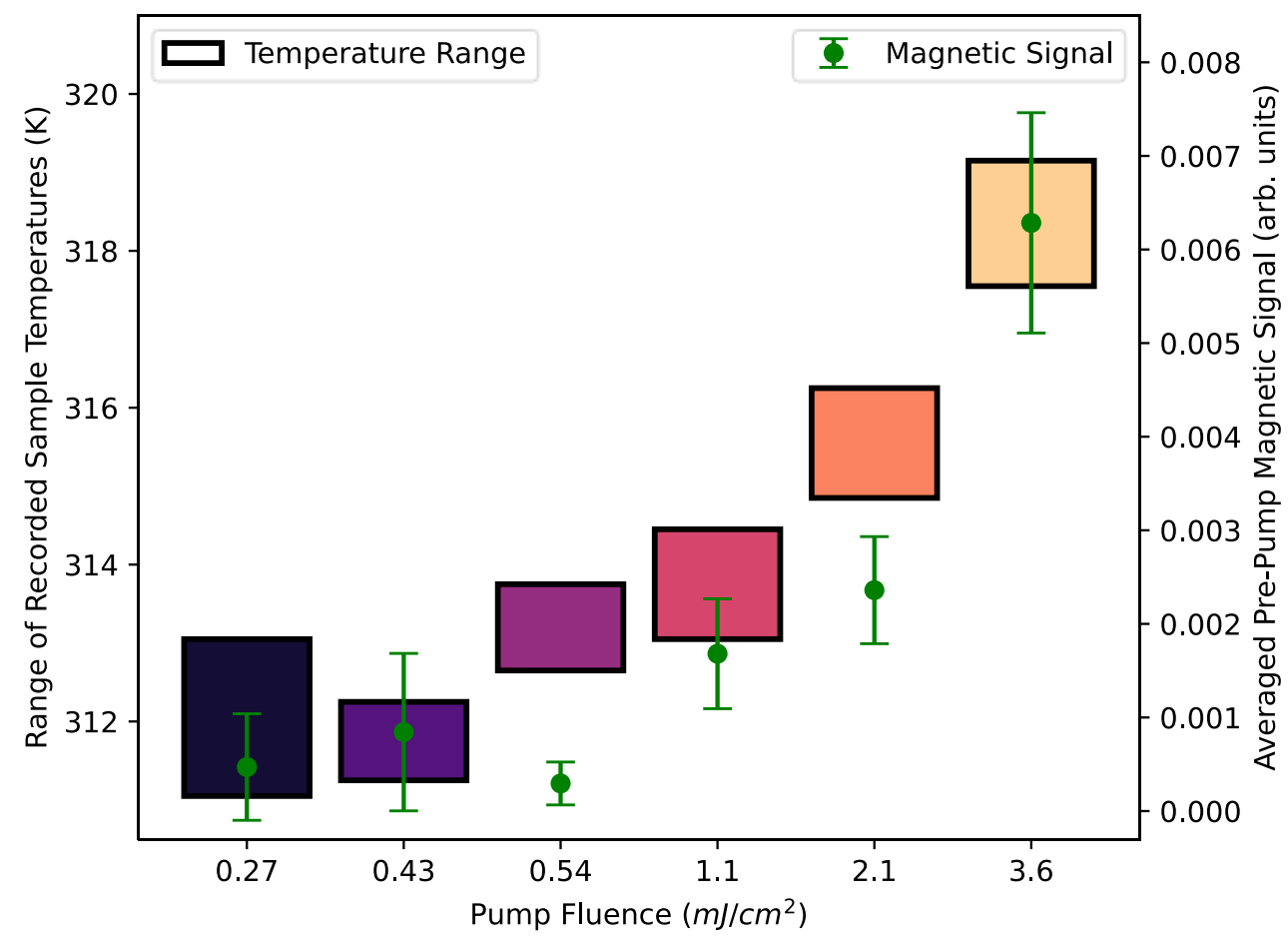

**Fig. S1. The relationship between static (pre-pump) magnetic signal and recorded sample temperature.** The magnetic asymmetry signals at a probe energy of 46.2 eV ± 0.2 eV (green data points with error bars) were calculated by averaging the pre-pump-excitation data ($t \leq 0$). As there were only 3- or 4-times points where $t \leq 0$, in the dynamic measurements (Fig. 2), a student t-value weighting was multiplied by the standard error to obtain the error bars shown. The temperature ranges (solid color bars) indicate the maximum and minimum temperatures recorded by the temperature probe during the scan.

### Section S2. Definition and Measurement of the Normalization Factor

The measurements in Fig. 1 and Fig. 2 took between 16 and 24 hrs of data collection per fluence to achieve the signal-to-noise ratio presented. One issue associated with taking data over these long timescales was sample damage induced by the pump and probe lasers. The sample was very small (few mm diameter) and, in many areas, the surface roughness was too large to make a measurement. Therefore, we were limited to taking data in one sample location per fluence. The pump pulses were low energy (≈ 1.59 eV), but high fluence (few $mJ/cm^2$ range) source of damage, while the probe laser has a comparatively much lower fluence (four orders of magnitude lower) but significantly higher photon energy (30-73 eV range). The sample damage reduced the magnetic signal over time. To prevent sample damage from influencing the measured dynamic behavior, the order in which the time points were measured was randomized. This random order was then repeated hundreds of times. This prevented systematic errors that would arise when measuring the time points sequentially. Nonetheless, the sample damage reduced the overall amplitude of the signal. The amount of damage differed between different pump fluences and sample locations.

After taking the original data with 6 fluences on 6 different sample locations, we devised a normalization procedure to allow us to compare the magnitudes of signals obtained despite sample inhomogeneities and fluence-dependent damage from the pump laser. We called this the normalization factor, denoted (NF). The NF is calculated in Table 1 and applied to all the data in Figs. 1 and Fig. 2. The steps used to determine the NF value are described below.

The measurement of the NF was made after the original 16-24 hr scans of varying fluence depicted in Figs. 1 and Fig. 2. The sample locations used for the original scans were already highly damaged. Therefore, we chose 6 new high reflectivity, undamaged locations on the sample for the NF measurement. These 6 sample locations were used to determine the NFs for the 6 fluences used in the study and appear in Table 1, column 1. The measurement of the NF consisted of two steps: the CV measurement and the MV measurement.

The first step is the control value (CV) measurement. To obtain the CV, we measured the magnetic signal with 0.54 $mJ/cm^2$ pumping, taken at a fixed pump-probe delay (30 ps). The fluence of 0.54 $mJ/cm^2$ was chosen because it induces a well-defined magnetic response without excessive sample damage. The time delay of 30 ps was chosen as this is where the signal is largest for 0.54 $mJ/cm^2$ pumping. By measuring each new sample location under the same conditions, we determined the severity of sample inhomogeneities and attempted to mitigate their influence. We calculated the mean of the six measurements and use this to normalize each CV measurement. The normalized CV values are depicted in column 2 of Table 1. The maximum deviation from the mean was 20%, i.e. CV = 1.2 (at location 2).

The second step is the measurement value (MV) measurement. The MV is obtained by measuring the TMOKE signal for each of the six fluences with a pump-probe delay chosen to maximize the signal. Each fluence is measured on a different one of the six sample locations from the CV measurement. The time delays for the MV measurements were chosen to maximize the signal for each specific fluence. The chosen time delays were: 20 ps, 30 ps, 30 ps, 70 ps, 200 ps, and -10 ps respectively. The MV values are listed in Table 1, column 4. Unlike the CV values, the MV values across different sample locations should not be similar to each other as they were taken at different fluences and different time delays.

Before both the CV and MV measurements, the sample had to thermalize with the relevant pump fluence so that the temperature of the sample would be stable throughout the scan. The probe was blocked while thermalizing with the pump (to limit sample damage from EUV). Furthermore, we kept each measurement as short as possible to limit sample damage. The timing was as follows: thermalize for 35 mins with 0.54 $mJ/cm^2$, measure the CV for 1 hour, thermalize for 15 mins with chosen fluence, measure the MV for 1 hour. These timings were a trade-off between improving signal-to-noise and minimizing exposure to the laser beams.

Finally, in Table 1, column 5, we recorded the measured asymmetries from the full dynamic fluence-dependent traces at the same specific fluences, and time delays used for the MV. These values are the real values (RVs). The RVs give us information on how much the measured signal was reduced due to sample damage over a long scan compared to the relatively short MV scans. By combining the values of the CV, MV and RV, the normalization factor can be calculated using the formula in Table 1, column 6.

For all fluences, except 0.27 $mJ/cm^2$, the normalization factor was greater than 1, i.e. the normalization process increased the magnitude of the signal. An NF>1 is expected in order to compensate for sample damage that occurs over the dynamic scans. The reason the 0.27 $mJ/cm^2$ fluence NF was not >1, may be because it had the least pumping, and therefore the least sample damage. Furthermore, the error bars on the 0.27 $mJ/cm^2$ NF were quite large due to the small overall magnetic signal induced by this fluence.

| Sample Location | Control Value CV | Fluence (mJ/cm$^2$) | Measured Value MV (1e-3) | Value in Real Data RV (1e-3) | Normalization Factor NF = (MV / RV) *1/CV |
|---|---|---|---|---|---|
| 1 | 0.96 ± 0.17 | 0.27 | 1.9 ± 0.9 | 2.5 ± 0.3 | 0.8 ± 0.4 |
| 2 | 1.20 ± 0.22 | 0.43 | 4.8 ± 1.1 | 2.9 ± 0.4 | 1.4 ± 0.4 |
| 3 | 1.00 ± 0.17 | 0.54 | 5.7 ± 0.9 | 4.1 ± 0.3 | 1.4 ± 0.3 |
| 4 | 0.96 ± 0.15 | 1.1 | 9.0 ± 0.8 | 5.2 ± 0.4 | 1.8 ± 0.3 |
| 5 | 0.98 ± 0.15 | 2.1 | 8.4 ± 0.9 | 4.7 ± 0.4 | 1.8 ± 0.4 |
| 6 | 0.90 ± 0.20 | 3.6 | 6.0 ± 1.0 | 5.8 ± 0.4 | 1.1 ± 0.3 |

**Table S1. Calculations of the normalization factors used for each fluence.** Error bars in the CV, MV, and RV represent the standard error based on repeated measurements. The errors in NF are calculated using standard error propagation techniques. Error on the NF is not represented in the error bars on figures in the main text.

## Section S3. Materials Information Tb166

Single crystals of $TbMn_6Sn_6$ were grown by self-flux methods as described by Jones *et al.* [5]. For completeness, we briefly summarize the origin of the spin reorientation in $TbMn_6Sn_6$. The spin reorientation in $TbMn_6Sn_6$ is a first-order phase transition: when the temperature is raised above $T_{\mathrm{SR}}$, the magnetic moments rotate from out-of-plane to in-plane. The ferrimagnetic order is governed by an indirect Mn–Mn exchange interaction mediated by antiparallel Tb moments. The transition originates from the competition between the easy-axis anisotropy of Tb and the easy-plane anisotropy of Mn. Upon heating toward the Ferrimagnetic Néel temperature ($T_{\mathrm{FN}}$), the Tb magnetic moment and its associated anisotropy decrease more rapidly than those of Mn because the molecular field acting on Tb (≈ 600 K) is much weaker than the strong in-plane Mn–Mn exchange (≈ 3000 K). As a result, around $T_{\mathrm{SR}} \approx (T_{\mathrm{FN}} - 100\text{ K})$, the Mn easy-plane anisotropy dominates, causing the net magnetization to rotate into the basal plane while maintaining the ferrimagnetic alignment between the Tb and Mn sublattices.

## Section S4. Model Equations

While the demagnetizing field, $\mathbf{H}_{\mathbf{demag}}(t)$, is much weaker than the anisotropy field, we still included an approximation of the demagnetizing field in the model. To calculate the demagnetizing field, we approximated the sample as a cylinder. The height of the sample was 0.7 mm and the diameter was 3.3 mm at its widest and 1.9 mm at its narrowest. We approximated this as a cylinder with a diameter four times its height. We then used an approximation [65] for the demagnetizing factors to obtain a demagnetizing field of:

$$\mathbf{H}_{\mathbf{demag}}(t) = \begin{pmatrix} -0.18\, M_x(t) \\ -0.18\, M_y(t) \\ -0.64\, M_z(t) \end{pmatrix} \tag{S1}$$

where $M_{x,y,z}$ are the components of the magnetization projected along different axes. These are calculated from projects of the instantaneous $\mathbf{m}(t)$ multiplied by the temperature-dependent saturation magnetization $M_{\mathrm{s}}$. The saturation magnetization depends on the temperature of the spin system, which in itself depends on time. Therefore, we write $M_{\mathrm{s}}$ as: $M_{\mathrm{s}}\left(T_{\mathrm{spin}}(t)\right)$.

The saturation magnetization is calculated based on the temperature-dependent Mn and Tb moments:

$$M_s(T_{spin}) = \left(6\mu_{Mn}(T_{spin}) - \mu_{Tb}(T_{spin})\right)/V. \qquad (S2)$$

The Tb moment per atom, $\mu_{Tb}(T_{spin})$, is subtracted due to its ferrimagnetic alignment with Mn. The Mn moment per atom, $\mu_{Mn}(T_{spin})$, is multiplied by 6 as there are 6 Mn atoms per unit cell. The temperature-dependent Tb and Mn magnetic moments come from neutron scattering data from El-Idrissi *et al.*(*16*) This total moment is then divided by the volume of the unit cell, $V$, to convert to total magnetization. $V$ is calculated based on the unit cell vectors given by Jones *et al.*[5], i.e. $V = a^2c$ where a = 5.538 Å and c = 9.0326 Å.

$\mathbf{H_{anis}}(t)$, can be calculated by taking the vector gradient of the anisotropy energy, dividing by the volume averaged magnetic moment, and considering only terms that can exert a torque. We calculate the anisotropy field, $\mathbf{H_{anis}}(t)$, using the element-specific, temperature-dependent, magnetic anisotropy energies, $MAE_{Tb}$ and $MAE_{Mn}$, as described by Jones *et al.* $\mathbf{H_{anis}}(t)$ drives the temperature-dependent dynamics and is given by the following expression:

$$\mathbf{H_{anis}}(t) = \begin{pmatrix} 0 \\ 0 \\ m_z(t)\left(\frac{MAE_{Tb}\left(T_{spin}(t)\right)}{\mu_0 M_s\left(T_{spin}(t)\right)} + \frac{6MAE_{Mn}\left(T_{spin}(t)\right)}{\mu_0 M_s\left(T_{spin}(t)\right)}\right) \end{pmatrix} \qquad (S3)$$

The factor of 6 in the expression for Mn accounts for the six Mn atoms per cell. A factor of 2 usually appears in the conversion between magnetic anisotropy energy and anisotropy field. However, we omit this term based on the definition of MAE used by Jones *et al.*(*4*).

**Section S5. Changing the Model Temperature Profile**

Since the exact shape of the time-dependent spin temperature profile is unknown, Figs. S3 (a)-(g) simulate different parameters for the applied temperature profile based on Eq. 4. The effect of changing the incident fluence is shown experimentally in Fig. S2(a), and in simulation in Fig. S2(b), as displayed in the main text, Fig. 3. The parameters tested include the ratio of the amplitudes of the fast and slow spin temperature decay, $A$ and $(1 - A)$ as well as the timescales themselves. Unless otherwise specified in the subfigure legends of Fig. S2, the parameters used were the same as in the main text: $T_{initial} = 270$ K, $A = 1/2$, $\tau_{rise} = 1$ ps, $\tau_{decay1} = 3$ ps, $\tau_{decay2} = 200$ ps, and $\alpha = 1$. Additionally, $T_{excite} = 35$ K. There is good agreement across a wide range of temperature model parameters shown in Fig. S2. In the simulations in Fig. 3, a damping of $\alpha$ = 1 was used. More physical values of $\alpha$ (e.g. 0.01 to 0.1), are depicted in Fig. S2(h). The overall envelope of the rise and decay is similar for each damping factor.

In Fig. S2(g) the laser-induced temperature change is modeled as a step-function. No exponential increases or decreases were applied, instead an instantaneous $\Delta T$ was introduced at $t$ = 0. The purpose of investigating the temperature step-function was to isolate the effects of the model-specific rise times and decays from the system's intrinsic spin-reorientation timescale. The data from subfigure (b) is overlaid on (g) as dotted lines. The maximum temperature changes induced by the $T_{excite}$ in subfigure (b) were used to set the $\Delta T$ values for the step-function model. In the case of $\Delta T$ = 15.7 K, a partial reorientation takes place, as seen in subfigure (g). In the case of $\Delta T$ = 23.9 K or 31.5 K, full reorientations occur.

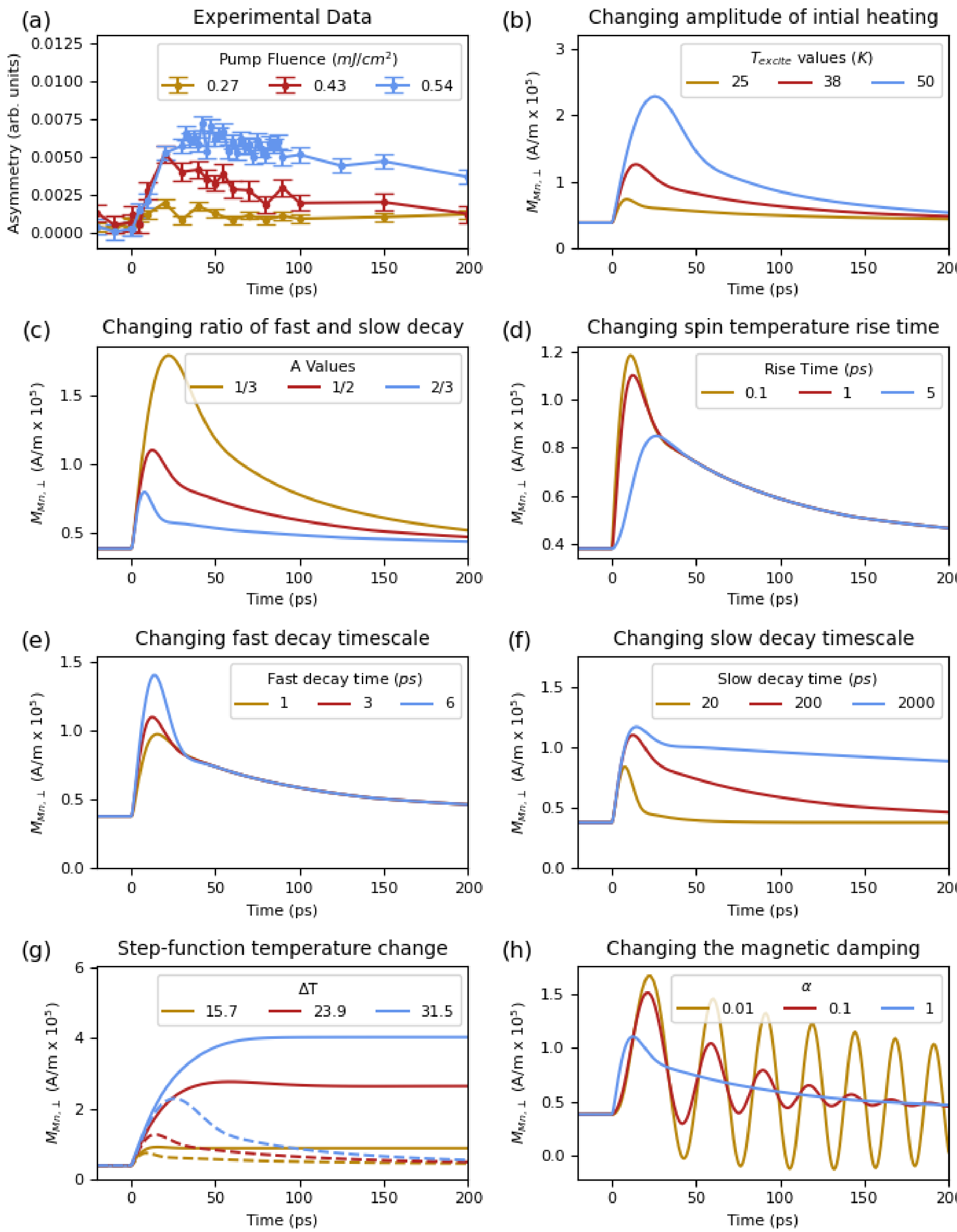

**Fig. S2. A low fluence dynamic model of Tb166.** The modeling is based on the temperature profile described by Eq. 4. Experimental data is presented for comparison in (a). The effects of changing the model parameters are shown. These are: (b) changing the amplitude of the initial heating $T_{\text{excite}}$; (c) changing the ratio of the amplitudes of the fast and slow temperature decays, $A$ and $(1 - A)$; (d) changing the rise time of the spin bath temperature, $\tau_{\text{rise}}$; (e) changing the fast decay timescale, $\tau_{\text{decay1}}$; and (g) changing the slow decay timescale, $\tau_{\text{decay2}}$. In (g), a step-function

temperature change is used instead of a three-exponential model. $\Delta T$ values are chosen to match the maximum temperature changes from subfigure (bb) and the data from subfigure (b) is overlaid as dotted lines. In (h), the magnetic damping factor, $\alpha$, is varied.

## Section S6. Modeling the High Fluence Data

For the highest fluence experimental data, 3.6 mJ/cm$^2$ and 2.1 mJ/cm$^2$, all measurements after $t$ = 0 are above $T_{\mathrm{SR}}$ so there are no reorientations. In this case, the LL or LLB equations are not required as there is no precessional motion to be modeled. The dynamics in the high fluence section are treated purely in terms of the change in $M_s$ as a function of changing sample temperature.

The experimental data from a fluence of 1.1 mJ/cm$^2$ is more difficult to model because, in this case, the sample passes through $T_{\mathrm{FN}}$ on early timescales then reaches $T_{\mathrm{SR}}$ on longer timescales. Therefore, measurements are in the regime of $T_{\mathrm{FN}}$ and will also require a precessional model for the near $T_{\mathrm{SR}}$ behavior. This data would be best modeled with the LLB equation and so we omit its treatment as it is beyond the scope of this work.

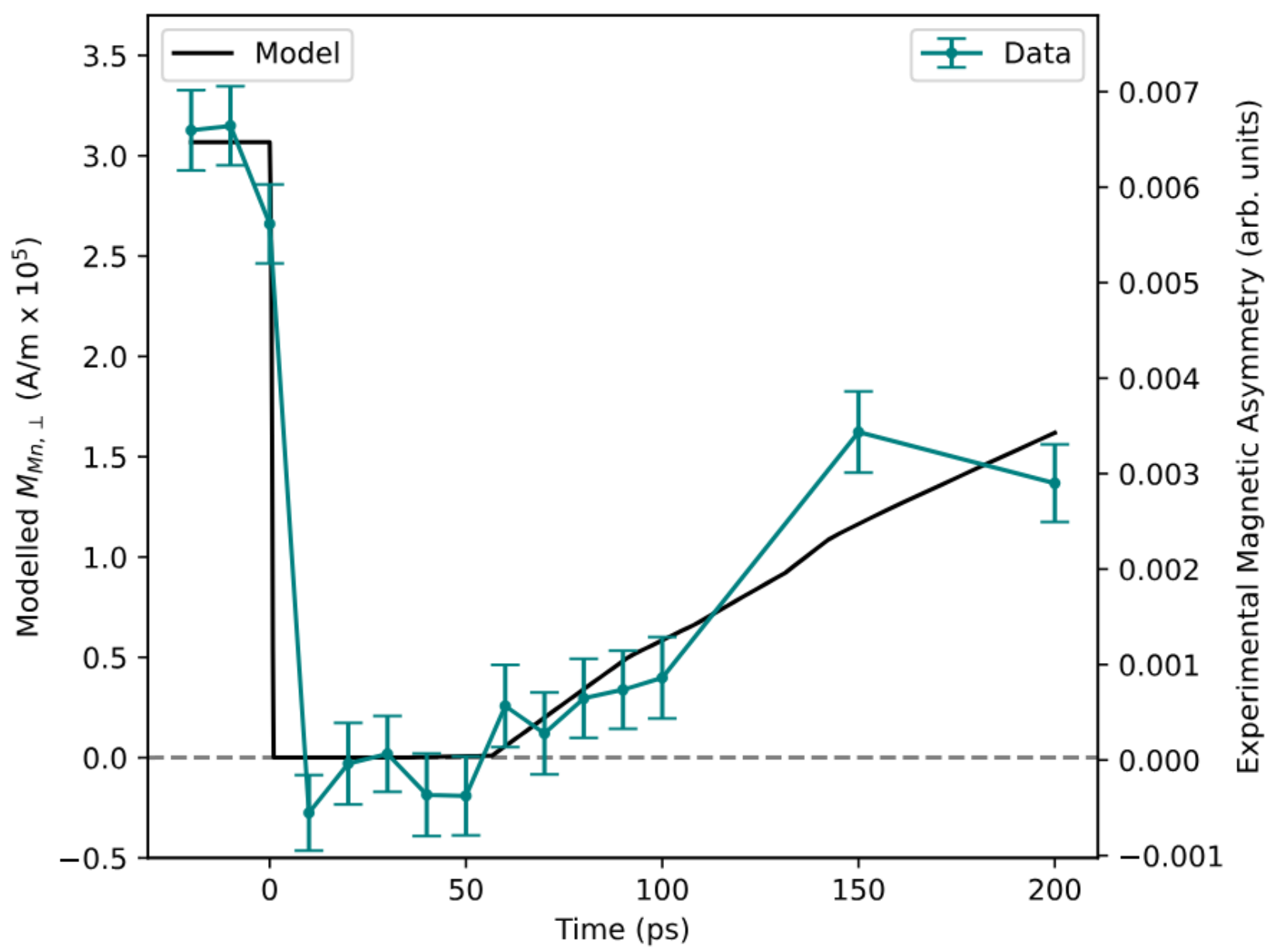

**Fig. S3. Modeling the high fluence behavior of Tb166.** The model utilizes an initial temperature of 294 K, a $\boldsymbol{T}_{\mathbf{excite}}$ amplitude of 275 K, and a slow decay time constant of 800 ps. The experimental data is for a fluence of 3.6 mJ/cm$^2$.

In Fig. S3, the experimental data for 3.6 mJ/cm$^2$ is plotted. A three-exponential temperature model is applied. For the model, a larger initial sample temperature (294 K) was used compared to the low fluence modeling (270 K) as the initial in-plane signal was larger for the high fluence data. The slow decay time was increased to 800 ps to match the experimental results, and the $A_1$ amplitude was increased to 275 K (note that 3.6 mJ/cm$^2$ is 6 times larger than the largest experimental fluence in the low fluence results). The increase in the slow decay timescale is understandable as there are many competing timescales for heat transport in the sample and it may be more difficult for a sample to dissipate additional heat from its spin system when the baseline temperature of the sample is hotter.